\documentclass[twocolumn,preprintnumbers, amssymb,amsmath,aps,floatfix,prl,nofootinbib,superscriptaddress,showpacs]{revtex4-1}

\usepackage{epsfig}
\usepackage{bm}
\usepackage{amssymb}
\usepackage{amsmath}
\usepackage{color}
\usepackage{subfigure}
\usepackage[colorlinks,
            linkcolor=blue,
            anchorcolor=black,
            citecolor=blue
            ]{hyperref}

\begin{document}
\title{Acoplanarity of Lepton Pair to Probe the Electromagnetic Property of Quark Matter}

\author{Spencer Klein}
\affiliation{Nuclear Science Division, Lawrence Berkeley National
Laboratory, Berkeley, CA 94720, USA}

\author{A. H. Mueller}
\affiliation{Department of Physics, Columbia University, New York, NY 10027, USA}

\author{Bo-Wen Xiao}
\affiliation{Key Laboratory of Quark and Lepton Physics (MOE) and Institute
of Particle Physics, Central China Normal University, Wuhan 430079, China}

\affiliation{Centre de Physique Th\'eorique, \'Ecole Polytechnique, 
CNRS, Universit\'e Paris-Saclay, Route de Saclay, 91128 Palaiseau, France.}

\author{Feng Yuan}
\affiliation{Nuclear Science Division, Lawrence Berkeley National
Laboratory, Berkeley, CA 94720, USA}

\begin{abstract}
We investigate the $P_T$-broadening effects in dilepton production through photon-photon scattering in heavy ion collisions. The QED multiple interaction effects with the medium is found to be consistent with a recent observation of low transverse momentum lepton pair from ATLAS collaboration at the LHC. We further comment on the magnetic effects and point out a number of ways to disentangle these two mechanisms. In particular, the rapidity dependence of the $P_T$-broadening effects provide a unique probe to the magnetic effects.  
\end{abstract}
\maketitle

{\it 1. Introduction.} Jet quenching is considered one of the major discoveries in relativistic heavy ion experiments from RHIC at Brookhaven National Laboratory and the LHC at CERN~\cite{Adcox:2001jp,Adler:2002xw,Aad:2010bu,Chatrchyan:2011sx}. These phenomena have been well formulated in QCD~\cite{Gyulassy:1993hr,Baier:1996kr,Baier:1996sk, Baier:1998kq, Zakharov:1996fv}, where the energy loss and $P_T$-broadening effects are closely related. The parameter $\hat q$ has been extracted from various experimental data, see, e.g., Ref.~\cite{Burke:2013yra}. Meanwhile, $\hat{q} L$ describes the typical transverse momentum squared that a parton acquires in the medium of length $L$. In the last few years, there have been significant progress in understanding the $P_T$-broadening effects in dijet, photon-jet, and hadron-jet productions in heavy ion collisions~\cite{Mueller:2016gko,{Mueller:2016xoc},{Chen:2016vem},{Chen:2016jfu},{Chen:2018fqu},Tannenbaum:2017afg,Adamczyk:2017yhe,Gyulassy:2018qhr}. At the LHC, the dominant broadening effect comes from the vacuum Sudakov effects~\cite{Mueller:2016gko} for the typical dijet kinematics~\cite{Aad:2010bu,Chatrchyan:2011sx}. On the other hand, the medium effect is comparable to the Sudakov effects at RHIC, and the STAR measurements have demonstrated the $P_T$-broadening effects in hadron-jet correlation~\cite{Adamczyk:2017yhe,Chen:2016vem}.  Future measurements at both LHC and RHIC should provide further information on the jet $P_T$-broadening physics.

More recently, both the ATLAS ~\cite{Aaboud:2018eph} collaboration at the LHC and the STAR \cite{Adam:2018tdm} collaboration at RHIC have found a new place to look for $P_T$-broadening effects - in the final states of dileptons that have been produced by the purely electromagnetic two-photon reaction: $\gamma\gamma\to\ell^+\ell^-$.  This reaction has been extensively studied in ultra-peripheral collisions (UPCs), where it is generally well described by lowest order quantum electrodynamics~\cite{Bertulani:1987tz,Baur:2007fv,Adams:2004rz,Bertulani:2005ru,Baltz:2007kq}. The lepton pairs have a very small pair $P_T$ (tens of ${\rm MeV}$), so the leptons are nearly coplanar.  However, ATLAS and STAR observed a significant $P_T$-broadening effects in dileptons from the reaction $\gamma\gamma\to\ell^+\ell^-$, in peripheral and (for ATLAS) central collisions.  This broadening hardens the STAR $p_\perp^2$ spectrum, and ATLAS sees a significant loss of coplanarity in moving from UPCs to central collisions. ATLAS also observes a small (order 1\% of the events) tail of events with high acoplanarity, even in UPCs~\cite{ATLAS:2016vdy}.  

In this paper, we study the mechanisms that can lead to this broadening. We extend our previous studies on the dijet azimuthal correlation to the di-lepton correlation and focus on two main areas. One is the QED Sudakov effect, where we show that the theory prediction for the UPC events agree very well with data from ATLAS~\cite{ATLAS:2016vdy}. Second, we investigate the medium effects, including the QED multiple interaction effects similar to the $P_T$-broadening of the QCD jet and the magnetic effects~\cite{Adam:2018tdm}. We also discuss how to disentangle these two mechanisms. 

The comparison of the $P_T$-broadening effects in QCD and QED is of crucial importance to understand the medium property in heavy ion collisions. The lepton's $P_T$-broadening effects is sensitive to the electromagnetic property of the quark-gluon plasma, whereas the jet $P_T$-broadening effects depends on the strong interaction property. The experimental and theoretical investigations of both phenomena will deepen our understanding of the hot medium created in these collisions. The clear measurements of lepton $P_T$-broadening effects from ATLAS and STAR~\cite{Aaboud:2018eph,{Adam:2018tdm}} should stimulate further study on dijet azimuthal correlations in heavy ion collisions. 

The rest of the paper is organized as follows. We first study the azimuthal angular correlation for dileptons in UPC in Sec.~2. Then, we investigate the medium effects, including the QED multiple scattering effects and the magnetic effects in Sec.~3 and 4, respectively. Finally, Sec. 5 summarizes the paper.

\begin{figure}
\includegraphics[width=8cm]{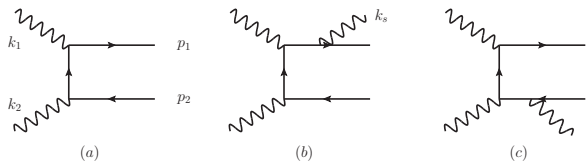}
\caption{The leading order and next-to-leading order QED Feynman diagrams for  lepton pair production through $\gamma\gamma$ processes: (a) the leading order diagram (interchange between $k_1$ and $k_2$ should be included as well); (b) soft photon radiation from the lepton; (c) soft photon radiation from the anti-lepton. Photon radiation from the lepton propagator is power suppressed.}
\label{fig1}
\end{figure}

{\it 2. Lepton Pair Production in Ultra Peripheral Heavy Ion Collisions}. The leading order production of lepton pair comes from photon-photon scattering, see, Fig.~\ref{fig1}(a). The outgoing leptons have momenta $p_1$ and $p_2$, individual transverse momenta $p_{1\perp}$ and $p_{2\perp}$, and rapidities $y_1$ and $y_2$, respectively. The leptons are produced dominantly back-to-back in the transverse plane, i.e., $|\vec{p}_\perp|=|\vec{p}_{1\perp}+\vec{p}_{2\perp}|\ll |p_{1\perp}|\sim  |p_{2\perp}|$.  The incoming photons have the following momenta, $k_1={P_{\perp}}/{\sqrt{s}}\left(e^{y_1}+e^{y_2}\right)P_A$ and $k_2={P_{\perp}}/{\sqrt{s}}\left(e^{-y_1}+e^{-y_2}\right)P_B$, where $P_\perp$ represents $|p_{1\perp}|\sim |p_{2\perp}|$, and the incoming nuclei have per-nucleon momenta $P_A$ and $P_B$. The differential cross section is conveniently written
\begin{eqnarray}
\frac{d\sigma(AB_{[\gamma\gamma]}\to\mu^+\mu^-)}{dy_1dy_2d^2p_{1\perp}d^2p_{2\perp}}&=&\sigma_0\int\frac{d^2r_\perp}{(2\pi)^2}e^{ip_\perp\cdot r_\perp}W(b_\perp;r_\perp) \ ,\nonumber
\end{eqnarray}
where $b_\perp$ denotes the centrality at a particular impact parameter of $AA$ collisions,  $\sigma_0=|\overline{\cal M}^{(0)}|^2/16\pi^2Q^4$ with $|\overline {\cal M}_0|^2=(4\pi)^2\alpha_e^2{2(t^2+u^2)}/{tu}$, $Q$ is the invariant mass for lepton pair, $t$ and $u$ are usual Mandelstam variables for the $2\to 2$ process. $W(b_\perp;r_\perp)$ contains incoming photon fluxes and all order Sudakov resummation,
\begin{eqnarray}
W(b_\perp;r_\perp)={\cal N}_{\gamma\gamma}(b_\perp;r_\perp)e^{-S_u(Q,m_\mu;r_\perp)}
\ ,\label{res}
\end{eqnarray}
where $S_u$ is the Sudakov factor and will be calculated below. Setting $S_u=0$ gets back to previous studies~\cite{Klein:2016yzr,Klein:2018cjh,Klein:1999qj,{Zha:2018ywo}}. The factor ${\cal N}_{\gamma\gamma}$ represents the incoming photon flux overlap,
\begin{eqnarray}
{\cal N}_{\gamma\gamma}(b_\perp;r_\perp)&=&x_ax_b\int  d^2k_{1\perp}d^2k_{2\perp}e^{i(k_{1\perp}+k_{2\perp})\cdot r_\perp}\nonumber\\
&&\times \left[f_A^\gamma(x_a,k_{1\perp})f_B^\gamma(x_a,k_{2\perp})\right]_b \ ,\label{photonflux}
\end{eqnarray}
where $x_a=k_1/P_A$, $x_b=k_2/P_B$. To simplify the above expression, we have introduced an impact parameter $b_\perp$-dependent photon flux: $\left[f_A^\gamma f_B^\gamma\right]_b=\int d^2b_{1\perp}d^2b_{2\perp}{\Theta(b_\perp)}N_\gamma(b_{1\perp},k_{1\perp})N_\gamma(b_{2\perp},k_{2\perp})$, where $\Theta(b)$ denotes the impact parameter constraints for a particular centrality with $\vec{b}_\perp=\vec{b}_{1\perp}-\vec{b}_{2\perp}$, and individual photon flux $N_\gamma(b_{1\perp},k_{1\perp})$ can be computed separately~\cite{Klein:2016yzr,Klein:2018cjh,Klein:1999qj,Baltz:2009jk,{Zha:2018ywo}}. Here, the interdependence between the impact parameter $b_{i\perp}$ and the photon's transverse momentum contribution $k_{i\perp}$ is ignored, which could introduce additional theoretical uncertainties. 

The Sudakov factor $S_u$ starts at one-loop order, where soft photon radiations contribute to the dominant logarithms in the kinematics of our interest. The typical Feynman diagrams for the real photon radiation are shown in Fig.~\ref{fig1}(b,c). Applying the Eikonal approximation, see, e.g.,~\cite{Sun:2015doa}, we obtain
\begin{equation}
{\cal M}^{(1)r}|^2_{soft}=e^2\frac{2p_1\cdot p_2}{p_1\cdot k_s p_2\cdot k_s}|{\cal M}^{(0)}|^2 \ .
\end{equation}
where ${\cal M}^{(0)}$ is the leading order Born amplitude and $k_s$ is the soft photon momentum. In the small total transverse momentum region $\ell_\perp\ll P_\perp$, we have the following behavior from the above contribution: $\frac{\alpha}{\pi^2}\frac{1}{\ell_\perp^2}\ln\frac{Q^2}{\ell_\perp^2+m_\mu^2}$, where $m_\mu$ is the lepton mass. In order to derive the one-loop result for $S_u$, we need to Fourier transform the above expression to the conjugate $r_\perp$-space, and add the virtual photon contributions. Because of the lepton mass $m_\mu$, the cancellation between the real and virtual diagrams will depend on the relative size of $\mu_r=c_0/r_\perp$ and $m_\mu$, where $c_0=2e^{-\gamma_E}$ with $\gamma_E$ the Euler's constant. In the end, we find at one-loop order~\cite{future},
\begin{equation}
S_{u}
=\begin{cases}
-\frac{\alpha}{2\pi} \ln^2\frac{Q^2}{\mu_r^2}\ ,  &\mu_r>m_\mu   \ , \\
-\frac{\alpha}{2\pi}\ln\frac{Q^2}{m_\mu^2}\left[\ln\frac{Q^2}{\mu_r^2}+\ln\frac{m_\mu^2}{\mu_r^2}\right]\ , &\mu_r<m_\mu \ .
\end{cases}
\label{su0}
\end{equation}
When the lepton mass is negligible, i.e., $\mu_r\gg m_\mu$, this leads to the same leading double logarithmic behavior as that in the back-to-back hadron production in $e^+e^-$ annihilation studied in Refs.~\cite{Collins:1981uk,{Collins:1981uw},{Collins:1981va}}. This provides an important cross check for our results. 

Combining the above Sudakov result with the incoming photon fluxes contributions in Eq.~(\ref{res}), we can calculate the total transverse momentum distribution. In order to simplify the numeric evaluation, we parameterize the $k_{i\perp}$ dependence for the incoming photon flux as simple Gaussian distributions with a typical width around $40\ \rm MeV$. The Gaussian width is also consistent with a fit to the STARlight \cite{Baltz:2009jk,Klein:2016yzr} simulation. In ATLAS, the azimuthal angular correlation of the lepton pair is studied: $\phi_\perp=\pi-(\phi_{1}-\phi_2)$ where $\phi_{1}$ and $\phi_2$ represent the azimuthal angles for the lepton and the anti-lepton, respectively. Figure~2 compares the different contributions to this correlation as function of the acoplanarity $\alpha=|\phi_\perp|/\pi$ at the LHC for the lepton pair production at mid-rapidity with lepton transverse momentum $P_\perp>4\ \rm GeV$ and invariant mass $10\ {\rm GeV}<M_{\mu\mu}<100\ {\rm GeV}$. The dotted line represents the primordial contribution from the two photon's transverse momenta, dashed curve for the perturbative one soft photon radiation and solid curve for the total contribution with resummation.  This result is in good agreement with the ATLAS UPC data~\cite{ATLAS:2016vdy}. In particular, the perturbative tail has been well described by the Sudakov formula. We have also checked that the so-called nucleus dissociation contribution (or incoherent nucleon contribution) is negligible in this kinematics because of additional $1/Z_A$ suppression. This provides an important baseline for the central collisions, which we discuss in the following sections. 

\begin{figure}
\includegraphics[width=7cm]{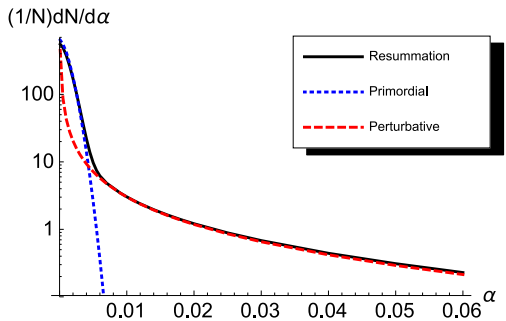}
\caption{Acoplanarity distribution for lepton pair production at mid-rapidity in UPC events at the LHC with a typical kinematics: lepton transverse momentum $P_\perp>4\ {\rm GeV}$ and pair invariant mass from 10 to 100 GeV. The detailed explanation of different curves is provided in the main text. The total contribution with resummation (solid curve) agrees well with the ATLAS measurement~\cite{ATLAS:2016vdy}.}
\label{upcalpha}
\end{figure}

{\it 3. Medium Effects in Central Collisions: Multiple Scattering.}  Both ATLAS and STAR focus on lepton pairs with small pair $P_T$, where the two photon scattering is the dominant channel in peripheral and even central heavy ion collisions~\cite{Aaboud:2018eph,Adam:2018tdm,Adam:2015gba}. The photon fluxes come from the charge distributions of incoming nuclei, and their contributions may not strongly depend on the centrality of the collisions. Therefore, in the following calculations, we assume that the total $P_T$ distribution from the incoming photons is the same for the peripheral and central collisions as in UPC events. 

In non-UPC heavy ion collisions, the ATLAS and STAR data show that the lepton pair have accumulated additional $P_T$-broadening. This could be from the interactions between the lepton pair and the medium. Because the leptons only carry electric charges, these interactions depend solely on the electromagnetic properties of the quark-gluon plasma (QGP) created in these collisions. 

The medium interactions are very much similar to the jet quenching and $P_T$-broadening mentioned in the Introduction. Like the QCD case, the leptons will suffer multiple scattering with the medium. To evaluate this contribution, we can follow the $P_T$-broadening calculations in QCD~\cite{Baier:1996vi,Baier:1996sk}. The multiple photon exchanges between the lepton and the medium can be formulated in a QED type time-ordered Wilson line
\begin{eqnarray}
\mathcal{U}_{\textrm{QED}} (x_\perp)&=&T\exp \left[-ie\int dz^- \int d^2 z_\perp G(x_\perp-z_\perp)\right.\nonumber\\
&&\times\left. \rho_e (z^-, z_\perp)\right],
\end{eqnarray}
where $\rho_e (z^-, z_\perp)$ is the electric charge source of the medium. The photon propagator $G(x_\perp)$ is defined as 
\begin{equation}
G(x_\perp)=\frac{1}{(2\pi)^2} \int d^2q_\perp \frac{1}{q_\perp^2+\lambda^2}e^{iq_\perp\cdot x_\perp}=\frac{1}{2\pi}\mathrm{K}_0(\lambda x_\perp)\ ,
\end{equation}
where $\lambda$ acts as an IR regulator similar to the Debye mass in QED.  
Analogous to the QCD $q\bar q$ dipole calculation, the QED multiple scattering amplitude between the $\ell^+\ell^-$ dipole with size $r_\perp$ and target medium can be written as
\begin{equation}
\langle\mathcal{U}_{\textrm{QED}} (b_\perp+\frac{1}{2}r_\perp) \mathcal{U}^{\dagger}_{\textrm{QED}} (b_\perp-\frac{1}{2}r_\perp)\rangle =\exp\left[-\frac{Q_{se}^2r_\perp^2}{4}\right], 
\end{equation}
where the analog of saturation momentum in QED $Q_{se}^2\equiv \frac{e^4}{4\pi}\ln \frac{1}{\lambda^2 r_\perp^2}\int dz^- \mu_e^2 (z^-)$. Here, $\mu_e^2$ is related to the local charge density fluctuations. The dipole size $r_\perp$ is large in the soft momentum transfer region, which makes $Q_{se}^2 r_\perp^2 \sim 1$. Therefore, we need to take into account the multiple scattering effects.

If we compare the above to the QCD dipole~\cite{Iancu:2002xk,Mueller:1999wm}, we will find the following differences. First, because the couplings in QED and QCD are dramatically different, this introduces a major difference for the $P_T$-broadening effects. Second, the saturation scales depend on the charge density. Since only quarks carry electric charge, the QED saturation scale will depend on the quark density, whereas the QCD saturation scale depends on both quark and gluon density. Their densities are proportional to the respective degree of freedoms if we assume the thermal distributions of the quarks and gluons: $\frac{21}{2}N_f: 16$~\cite{Blaizot:2014jna}. Here $N_f$ is the number of active flavors. After accounting for the color factor differences in the multiple scattering, we estimate the ratio between the QED and QCD saturation scales as 
\begin{equation}
\frac{\langle \hat q_{QED}L\rangle }{\langle\hat q_{QCD}L\rangle}=\frac{\alpha_e^2}{\alpha_s^2}\frac{\frac{21}{2}N_f\frac{2}{9}}{\frac{21}{2}N_f\frac{2}{9}+16\frac{1}{2}}= \frac{\alpha_e^2}{\alpha_s^2}\times\frac{7}{15}\ ,
\end{equation}
for $N_f=3$ and for quark jet, where $\langle \hat q L\rangle$ represents the saturation scale in the dipole formalism. For gluon jet, there is a factor of $C_A/C_F$. A few comments are in order. First, we assume that quark and gluons are thermalized at the same time, which may not be true~\cite{Blaizot:2014jna}. Second, we did not take into account the detailed effects from the medium property, such as the associated Debye masses for QED and QCD. In addition, for the QCD case, there is length dependent double logarithms~\cite{Liou:2013qya}. If this is to be taken into account, the above simple formula will not apply. Nevertheless, the above can serve as a simple formula for a rough estimate. 

\begin{figure}
\includegraphics[width=6cm]{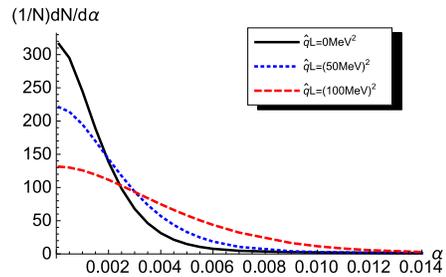}
\caption{Medium modifications to the acoplanarity distribution, with different values of the effective $\hat qL$.}
\label{central}
\end{figure}

If we assume the multiple scattering limit, we can modify the above $W(b_\perp;r_\perp)$ of Eq.~(\ref{res}) as,
\begin{eqnarray}
{\cal N}_{\gamma\gamma}(b_\perp;r_\perp)e^{-S_u(Q,m_\mu;r_\perp)}
e^{-\frac{\langle\hat q_{QED}L\rangle r_\perp^2}{4}}\ ,
\end{eqnarray}
where the last factor comes from the medium contribution to the di-lepton $P_T$-broadening effects. In Fig.~\ref{central}, we show this effects by imposing two different values of the $\hat qL$. Comparing these curves to the ATLAS measurements, we conclude that the effective $\langle \hat q_{QED}L\rangle$ range from $(100\ {\rm MeV})^2$ in most central collisions to $(50\ {\rm MeV})^2$ in non-central collisions. We can also estimate the QED energy loss~\cite{Baier:1996vi}. However, it is too small (few percent of $P_T$-broadening value) to have any observational effects. 

{\it 4. Medium Effects: Magnetic Fields.} There has been a suggestion that the $P_T$-broadening could come from the magnetic effects of the medium~\cite{Adam:2018tdm}, as a result of the Lorentz force: $\vec{B}\times \vec{V}$, where $\vec{B}$ and $\vec{V}$ are the magnetic field vector and the lepton's velocity, respectively.  The lepton bending is strongly correlated with the directions of the magnetic field and the lepton's momentum. If we can measure these correlations, we will be able to disentangle these mechanisms.

The initial magnetic fields generated by the colliding nuclei will contribute to an additional $P_T$-broadening effects. However, this effects is completely cancelled out by the effects from the electric fields in the leading power of $q_\perp/P_\perp$~\cite{Ivanov:1998ka,future}. This cancellation is also consistent with a factorization argument that the final state interaction effects vanishes in this process because of the opposite charges of the lepton pair.  

Some theorists have suggested that there is a residual magnetic field in the quark-gluon plasma after the collisions~\cite{Kharzeev:2009pj,Asakawa:2010bu,Skokov:2016yrj}. Because of the collision symmetry, the magnetic field only contains the perpendicular component $\vec{B}_\perp$. It has a nontrivial dependence on the impact parameter: increases from UPC to peripheral collisions but decreases toward more central collisions~\cite{Kharzeev:2009pj,Asakawa:2010bu,Skokov:2016yrj}. The ATLAS data does not appear to follow this trend. 

This is very different from the multiple interaction effects discussed above, which increases monotonically with the centrality. Furthermore, because the Lorentz force vanishes along the direction of the magnetic field, the $P_T$-broadening effects from the magnetic effects will have a non-trivial correlation with the event plane, which is correlated with the direction of the magnetic field~\cite{Kharzeev:2009pj,Asakawa:2010bu,Skokov:2016yrj}. 

\begin{figure}
\includegraphics[width=6cm]{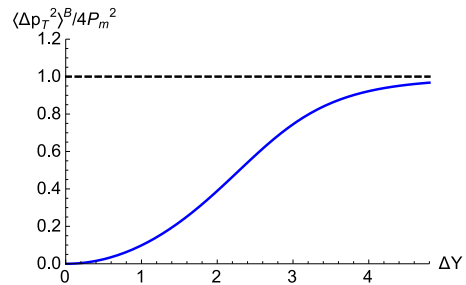}
\caption{Normalized magnetic effects on the $P_T$-broadening for the lepton pair as function of their rapidity difference $\Delta Y=|y_{\mu^+}-y_{\mu^-}|$ with $|y_\mu|<2.4$. }
\label{magneticdy}
\end{figure}

More importantly, the magnetic effects depends on the longitudinal velocity $v_z$ of the leptons. Therefore, if the lepton and the anti-lepton move in the same $z$ direction, the magnetic effects will cancel out in the total pair $P_T$. Because of the linear dependence on $v_z$, the total $P_T$-broadening effects for the pair can be formulated as
\begin{equation}
\langle \Delta p_\perp^2\rangle_{\mu^+\mu^-}^{B}=\langle {\rm \bf P_m^2}(b_\perp)\rangle \left[\tanh (y_+)-\tanh(y_-)\right]^2 \ ,
\end{equation}
where $ \langle {\rm \bf P_m^2}(b_\perp)\rangle$ represents the average $P_T$-broadening depending on the centrality of the collisions, $y_+$ and $y_-$ are rapidities for the lepton and the anti-lepton, respectively. Figure~\ref{magneticdy}, shows the normalized contribution as function $\Delta Y=|y_{\mu^+}-y_{\mu^-}|$ for a typical lepton transverse momentum $P_\perp=6\ {\rm GeV}$. As expected, the magnetic effects on the $P_T$-broadening increases with $\Delta Y$. On the other hand, the multiple scattering effects discussed in the last section depends on the charge density of the medium and will not change in this rapidity range. Therefore, the difference between the $P_T$-broadening effects at different $\Delta Y$ can be used as an effective measure to the magnetic effects:
\begin{equation}
    \left[\langle\Delta p_\perp^2\rangle_{\Delta Y=3}-
        \langle\Delta p_\perp^2\rangle_{\Delta Y=0}\right]_{b_\perp}\propto \langle \vec{B}_\perp^2\rangle_{b_\perp} \ ,
\end{equation}
which will depend on the centrality of heavy ion collisions. This will help to investigate other magnetic effects in heavy ion collisions, such as the chiral magnetic effects~\cite{Kharzeev:2009pj,Skokov:2016yrj}.

{\it Summary and Discussions.}
We have investigated the di-lepton production at very low total transverse momentum in heavy ion collisions to probe the electromagnetic property of the quark-gluon plasma. In the theoretical calculations, we take into account two important contributions: one is the soft photon radiation with Sudakov resummation; one is the medium interaction leading to $P_T$-broadening for the leptons when they traverse through the medium. By including a Sudakov resummation, we have shown that the theory predictions can well describe the azimuthal angular correlation of the lepton pair  in the UPC events. 

We have also shown that the $P_T$-broadening effects found by the ATLAS collaboration can be described by the multiple scattering of the leptons in the medium, where the effective $\langle \hat q_{QED}L\rangle$ of order of $(50\ {\rm MeV})^2$ to $(100\ {\rm MeV})^2$ are in agreement with a parametric estimate of the QED and QCD effects of the quark-gluon plasma. We have investigated the $P_T$-broadening effects from the magnetic fields as well, and pointed out there are a number of ways to distinguish these two mechanisms, through a detailed study on: (1) the centrality dependence of the effects; (2) the correlation with the magnetic field (or reaction plane); (3) the rapidity dependence for the lepton pair. We emphasized that the magnetic effects depends on the rapidity difference between the lepton and the anti-lepton. This dependence can be used to determine the strength of the magnetic field. 

In summary, our study demonstrated that the azimuthal correlation of the lepton pair in low total transverse momentum region is a powerful tool to investigate the electromagnetic property of the quark-gluon plasma in heavy ion collisions. This shall stimulate further experimental and theoretical studies. In particular, we hope the rapidity dependence of the $P_T$-broadening effects can be measured to uniquely probe the magnetic effects.

\begin{acknowledgments}
We are grateful to Aaron Angerami, Brian Cole, Peter Jacobs, Volker Koch, Derek Teaney, Zhangbu Xu for interesting discussions and comments. This material is based upon work partially supported by the Natural Science Foundation of China (NSFC) under Grant Nos.~11575070, the U.S. Department of Energy, Office of Science, Office of Nuclear Physics, under contract number DE-AC02-05CH11231.
\end{acknowledgments}

\end{document}